\documentclass[conference]{IEEEtran}
\IEEEoverridecommandlockouts

\usepackage{cite}
\usepackage{amsmath,amssymb,amsfonts}
\usepackage{algorithmic}
\usepackage{graphicx}
\usepackage{textcomp}
\usepackage{hyperref}

\usepackage{xcolor}
\def\BibTeX{{\rm B\kern-.05em{\sc i\kern-.025em b}\kern-.08em
    T\kern-.1667em\lower.7ex\hbox{E}\kern-.125emX}}

\begin{document}
\title{Predicting power grid frequency dynamics with invertible Koopman-based architectures
\thanks{Corresponding author: Xiao Li (lixiao000@yeah.net) and Benjamin Schäfer (benjamin.schaefer@kit.edu). This research is funded by Sino-German (CSC-DAAD) Postdoc Scholarship Program(91870333) and the Helmholtz Association's Initiative and Networking Fund through Helmholtz AI under grant no. VH-NG-1727.}
}

\author{\IEEEauthorblockN{Eric Lupascu}
\IEEEauthorblockA{\parbox{5cm}{\centering\textit{Institute of Automation and Applied Informatics}} \\
\textit{Karlsruhe Institute of Technology}\\
Eggenstein-Leopoldshafen, Germany \\
uiqte@student.kit.edu}
\and
\IEEEauthorblockN{Xiao Li\textsuperscript{*}}
\IEEEauthorblockA{\parbox{5cm}{\centering\textit{Institute of Automation and Applied Informatics}} \\
\textit{Karlsruhe Institute of Technology}\\
Eggenstein-Leopoldshafen, Germany \\
lixiao000@yeah.net}
\and
\IEEEauthorblockN{Benjamin Schäfer\textsuperscript{*}}
\IEEEauthorblockA{\parbox{5cm}{\centering\textit{Institute of Automation and Applied Informatics}} \\
\textit{Karlsruhe Institute of Technology}\\
Eggenstein-Leopoldshafen, Germany \\
benjamin.schaefer@kit.edu}
}

\maketitle

\begin{abstract}
  The system frequency is a critical measure of power system stability and understanding, and modeling it are key to ensure reliable power system operations. Koopman-based autoencoders are effective at approximating complex nonlinear data patterns, with potential applications in the frequency dynamics of power systems. However, their non-invertibility can result in a distorted latent representation, leading to significant prediction errors. Invertible neural networks (INNs) in combination with the Koopman operator framework provide a promising approach to address these limitations. In this study, we analyze different INN architectures and train them on simulation datasets. We further apply extensions to the networks to address inherent limitations of INNs and evaluate their impact. We find that coupling-layer INNs achieve the best performance when used in isolation. In addition, we demonstrate that hybrid approaches can improve the performance when combined with suitable INNs, while reducing the generalization capabilities in combination with disadvantageous architectures. Overall, our results provide a clearer overview of how architectural choices influence INN performance, offering guidance for selecting and designing INNs for modeling power system frequency dynamics.
\end{abstract}

\begin{IEEEkeywords}
Frequency oscillation, Koopman operator, Invertible neural network, Power system
\end{IEEEkeywords}

\section{Introduction}
The electric power grid is one of the most critical infrastructures of modern society, supporting essential sectors such as healthcare, industry, communication, and public services. A central challenge in power system operation is ensuring system stability, which means that the grid can continue to operate safely and effectively after disturbances or sudden changes in load. The increasing integration of renewable energy sources, which are often inverter-based  and provide little inherent inertia \cite{b1}, makes maintaining this stability increasingly challenging, especially the frequency stability \cite{b2}.

Therefore, predicting how these states evolve over time is crucial. One widely used method for obtaining a linear approximation of system dynamics directly from data is Koopman operator-based modeling (see, for example, \cite{b3}). One implementation of this method is  dynamic mode decomposition (DMD), which finds the best-fitting linear operator that describes the evolution of the systems dynamic from state at time $k$ to the state at time $k+1$. Since this linear operator acts globally on all system states, it cannot capture nonlinear dynamics, which require a state-dependent operator. With the increasing application of neural networks, another approach emerged. Instead of fitting a linear operator directly to the nonlinear dynamics, we use autoencoders to learn an observable space in which the dynamics can be approximated linearly. The linear operator is applied in this space, and the prediction is then mapped back to the original state space, as demonstrated in \cite{b4,b5}. These approaches are purely data-driven and have significant advantages over traditional model-driven simulation methods, particularly when models are unavailable or unreliable.

However, the usefulness of Koopman-based models critically depends on the quality of the learned observables. Since Koopman analysis decomposes the dynamics into eigenvalues, eigenfunctions, and Koopman modes, any loss of information in the mapping from the physical state space to the observable space directly affects the accuracy of the recovered spectral quantities. This motivates the use of INNs, which enforce a bijective mapping between state space and latent space by design. Because no information is lost, the latent representation supports a more reliable approximation of the underlying dynamics and enables a robust estimation of Koopman spectral components. Recent literature therefore increasingly combines INN architectures with Koopman-operator--based modeling to improve both predictive performance and the interpretability of learned dynamical structures.

While INNs provide bijective representations by design, they can be constructed using markedly different architectural strategies, each associated with different properties. Examples include coupling-based flows, residual flows and neural ODEs. Nevertheless, INNs have their own limitations. Their expressiveness is significantly lower than that of conventional neural networks, and the requirement for equal input and output dimensionality conflicts with the Koopman operator's need for dimensional lifting, which embeds low-dimensional non-linear dynamics in higher-dimensional linear spaces \cite{b6}. One possible solution for this problem is the Augmented Invertible Koopman Autoencoder (AIKAE) proposed by Frion et al. \cite{b7}. As shown in Fig. \ref{fig:hybrid_INN}, it parallels the extension network and the INN. This mitigates the dimensionality limitations of standard INNs by introducing additional latent augmentation variables, enabling effective dimension lifting while preserving invertibility. 

To further explore these architectural considerations, this study focuses on the following two key research questions:

\begin{enumerate}
    \item How does the architectural choice of the INN influence the overall performance of our model?
    \item How can we combine invertible and non-invertible layers to achieve better performance?
\end{enumerate}

The remainder of this paper is organized as follows. In Chapter \ref{sec:architecture} we look deeper into the architectural strategies of INNs. Chapter \ref{sec:exp} shows our experiments done regarding base INNs \ref{subsec:base} and hybrid models \ref{subsec:hybrid} before we discuss our results in chapter \ref{sec:Disc} and finally draw our conclusion.


\section{Architecture of INNs}
\label{sec:architecture}

INNs are characterized by bijective and differentiable mappings that enable efficient forward and inverse computations, as well as tractable Jacobians. These properties ensure a lossless information flow between inputs and latent representations. 

\subsection{Categories of INNs}
\subsubsection{Coupling Layers}

The most common strategy for INNs is the use of coupling layers. A general coupling layer partitions the input vector into two disjoint sub-vectors $(x_A, x_B)$ and defines a bijective transformation $g: \mathbb{R}^D \to \mathbb{R}^D$ such that the transformation for the output sub-vectors $(y_A, y_B)$ is given by $y_A = x_A$ and $y_B = h(x_B; Q(x_A))$. Here, $h$ is an invertible coupling function parameterized by $Q(x_A)$, with $Q$ being an arbitrary conditioner function that depends only on $x_A$. This construction leads to a block-triangular Jacobian, which simplifies determinant computation and ensures easy inversion using the inverse coupling function $h^{-1}$.

A simple special case is the additive coupling layer, where $h(x_B; Q(x_A)) = x_B + Q(x_A)$. This layer is trivially invertible via subtraction and has a unit Jacobian determinant, making it computationally attractive (applied in \textbf{NICE} \cite{b8}).

However, the limited expressiveness of additive coupling motivated the development of more flexible layers, called affine coupling layers. An affine coupling layer extends the additive case by introducing learned scale and translation functions:
\begin{equation}
\begin{aligned}
y_A &= x_A,\qquad \\
y_B &= x_B \odot \exp(Q_s(x_A)) + Q_t(x_A),
\end{aligned}
\end{equation}
The inverse is obtained by corresponding subtraction and rescaling. Because the Jacobian remains block-triangular, the determinant is easy to compute and independent of derivatives of $Q_s$ or $Q_t$, allowing these subnetworks to be arbitrarily expressive.

To ensure that all dimensions are eventually transformed, successive coupling layers alternate the input partition. Stacking such layers yields the class of Coupling Flow INNs (CF-INNs). Affine coupling layers form the core of influential flow architectures, including \textbf{RealNVP}~\cite{b9} and \textbf{Glow}~\cite{b10}. Nevertheless, even with alternating partitions, only transforming half of the input limits the expressiveness of individual blocks. Hierarchical Invertible Neural Transport (\textbf{HINT})~\cite{b11} addresses this constraint by introducing a multi-level, recursive partitioning of the input space.

\subsubsection{Residual blocks}

Residual Flows are constructed through the composition of functions of the form $g(x) = x + F(x)$, known as residual connections with a residual block $F$.

Such residual connections were first used for reversible neural networks in i-RevNet~\cite{b12} and RevNet~\cite{b13}. These networks split the input, similar to additive coupling layers, and transform the vector $x = (x_A, x_B)$ by calculating $y_A = x_A + F(x_B)$ and subsequently $y_B = x_B + G(y_A)$. While this transformation is easily invertible, computing the Jacobian matrix is challenging. Moreover, these structures also rely on dimension partitioning.

A different approach is proposed by Behrmann et al.~\cite{b14} in the \textbf{iResNet}, who showed that residual connections are invertible if the Lipschitz constant is constrained to be less than one. In practice, this is enforced via spectral normalization of the weight matrices, and the inverse is approximated using a fixed-point iteration.
However, restricting the Lipschitz constant reduces expressiveness and the iResNet is not suited for memory-efficient applications \cite{b15}.

In the context of Koopman operator theory, INNs based on residual flow structures have been used, for example in~\cite{b16} and~\cite{b17}, where they provide invertible and expressive models for learning linear dynamics in latent space.

\subsubsection{\textbf{ODE-based} approaches}
Another class of invertible neural architectures is based on Neural Ordinary Differential Equations (ODEs) \cite{b18, b19}. This approach interprets residual connections as the discretization of first-order ODEs, thereby replacing discrete layers with a continuous transformation solved via a black-box ODE solver. While memory-efficient and expressive, continuous-time Neural ODEs are inherently unsuitable for Koopman operator learning, which requires discrete-time mappings from $t$ to $t+1$.

\subsection{Hybrid INNs}

\begin{figure}[htbp]
\centering
\includegraphics[width=\linewidth]{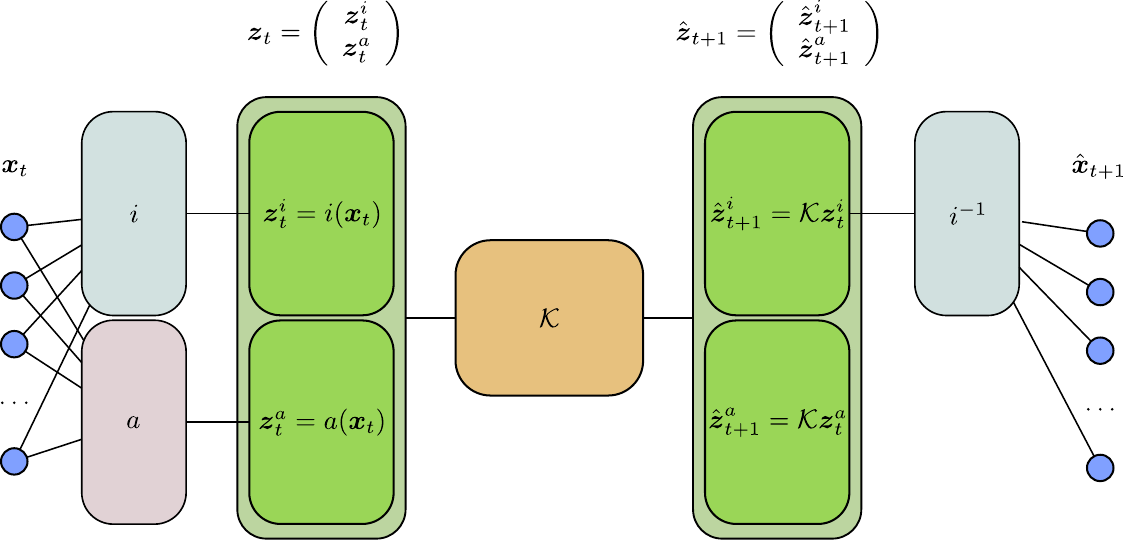}
\caption{Schematic representation of a hybrid INN inspired by the architecture introduced by \cite{b7}. INN $\boldsymbol{i}$ and a non-invertible extension network $\boldsymbol{a}$ are paralleled as an encoder, but only INN $\boldsymbol{i}$ is also used as a decoder. Koopman operator $\mathcal{K}$ is in the middle of the encoder and decoder.}
\label{fig:hybrid_INN}
\end{figure}

Maintaining the invertibility property requires that the latent representation cannot have a reduced dimensionality. Consequently, less relevant features must be preserved in the feature space. However, it is often desirable to augment or modify the latent space to improve the expressiveness and performance of the model, which can conflict with the invertibility constraints.
Previous works attempted to address this issue by appending zero padding to the state vector~\cite{b17,b20}, but this forces the model to reconstruct the artificial zeros and breaks the exact mathematical transformations needed for probability computations.

Frion et al. \cite{b7} proposes another approach (see Fig. \ref{fig:hybrid_INN}). 
In order to obtain additional features for a more expressive approximation of the Koopman operator, 
the input is processed not only by an invertible network $\boldsymbol{i}$ but also by a non-invertible network $\boldsymbol{a}$. 
The outputs of the two networks, $\boldsymbol{z}_t^{i}$ with dimension $p$ and $\boldsymbol{z}_t^{a}$ with dimension $m$, 
are concatenated into a single vector $z_t$ in latent space which is then passed to the Koopman operator. 
The Koopman operator, now defined on this higher-dimensional space, performs the state evolution on the concatenated vector. 
The Koopman prediction $\hat{\boldsymbol{z}}_{t+1}$ is subsequently split back into its two components 
$\hat{\boldsymbol{z}}_{t+1}^i$ and $\hat{\boldsymbol{z}}_{t+1}^a$, 
where only the invertible part $\hat{\boldsymbol{z}}_{t+1}^i$ of dimension $p$ is used to reconstruct the original state $\boldsymbol{x}_{t+1}$. 
A key advantage of this design is that the extension network $\boldsymbol{a}$ is free of INN structural constraints, allowing any neural network architecture and the use of downsampling or dimensionality reduction to efficiently generate additional features. 
The overall invertibility of the model is not compromised, 
since only the invertible part of the concatenated feature vector is used for reconstruction.

In this work, we extend this idea by evaluating this architecture not only with various base INN architectures, which has already been evaluated individually, but also with different types of extension layers. The goal is to gain further insight into which combination of INN and extension yields the best performance.
For the extensions, we employ three distinct types, each designed to capture complementary features and enhance the representation power of the INNs for Koopman operator approximation. First, we introduce a CNN-based extension, which leverages multi-scale convolutional kernels to extract local structures from the input data and to capture temporal patterns. Second, we use a kernel-based extension, which maps the input into high-dimensional feature spaces using nonlinear kernels, thereby modeling complex relationships and emphasizing similarity structures in the data; specifically, we employ the radial basis function (RBF) kernel. Finally, we apply a deep residual extension, also referred to as the multitimescale extension, which consists of a fully connected residual network with SiLU activations and skip connections to capture long-term dependencies and integrate frequency components across multiple scales.


\section{Experiments}
\label{sec:exp}
To properly evaluate the reviewed structures, we integrate the different INN architectures as well as the hybrid INNs into the Higher-Order Dynamic Mode Decomposition (HODMD) pipeline provided by \cite{b21} and assess their performance on two benchmark systems: the IEEE-14 bus system and the WECC-179 bus system. The IEEE system exhibits faster-decaying dynamics, whereas the WECC system shows weaker damping and significant inter-area oscillations. Dynamic trajectories are generated using the ANDES Python package \cite{b22} by applying temporary short-circuit faults at selected buses, with each fault lasting 0.1 seconds. For the IEEE system, 11 trajectories are produced, of which 9 are used for training and 2 for testing, while the WECC system yields 99 trajectories, with 90 allocated for training and 9 for testing. 
Besides the ODE-based network, all INN architectures used in this work were implemented using the FrEIA package \cite{b23}, leveraging its modular design to construct each network from reusable coupling and transformation blocks.
The models are trained by jointly optimizing prediction accuracy and latent-space consistency. The prediction loss measures the discrepancy between the predicted and true trajectories, the Koopman loss enforces consistency of the latent-space representation under the learned Koopman operator. Prediction and Koopman losses are logarithmically scaled to improve numerical stability. Hyperparameters such as network depth, hidden sizes, time delay, learning rate, batch size, and activation functions are optimized for each INN architecture, while model-level parameters are fixed across architectures to preserve comparability. Validation losses on unseen trajectories are used to guide hyperparameter selection and ensure generalization. 

The code we used for the experiments can be found online\footnote{\href{https://github.com/KIT-IAI-DRACOS/inn-koopman-freqdyn-powergrid} {https://github.com/KIT-IAI-DRACOS/inn-koopman-freqdyn-powergrid}}. It provides a modular HODMD pipeline in which any INN architecture can be integrated as a drop-in component and augmented with arbitrary non-invertible networks. Additionally, scripts for generating power system fault trajectories for both benchmark systems are included.

\subsection{Base INNs}
\label{subsec:base}

As a first step, we evaluate the performance of the INNs without any extensions within the Koopman pipeline. 
The recommended metric for assessing HODMD-based models is the relative root mean squared error (RRMSE) \cite{b24}, defined as
\[
\text{RRMSE} =
\sqrt{
\frac{
\sum_{i=1}^{N_\text{traj}} \left\| x_{\text{true}}^{(i)} - x_{\text{pred}}^{(i)} \right\|^{2}
}{
\sum_{i=1}^{N_\text{traj}} \left\| x_{\text{true}}^{(i)} \right\|^{2}
}
},
\]
where $N_\text{traj}$ denotes the number of trajectories, $x_{\text{true}}$ the ground truth, and $x_{\text{pred}}$ the model prediction.

\begin{figure}[htbp]
\centering
\begin{minipage}{0.48\linewidth}
    \centering
    \includegraphics[width=\linewidth]{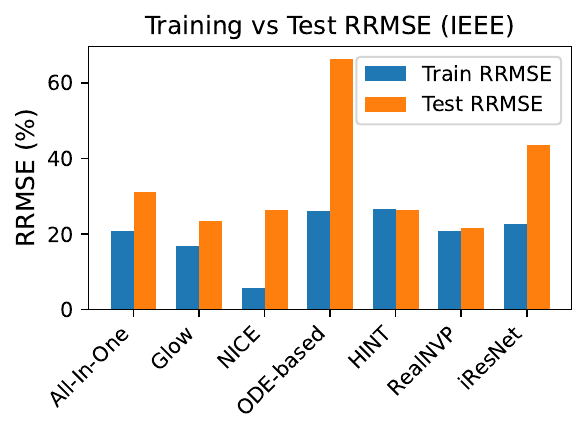}
\end{minipage}
\hfill
\begin{minipage}{0.48\linewidth}
    \centering
    \includegraphics[width=\linewidth]{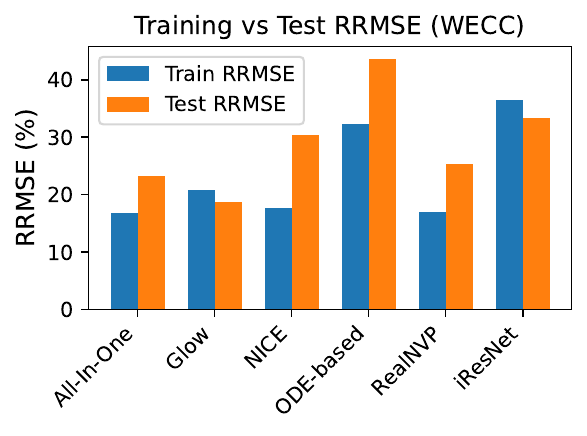}
\end{minipage}
\caption{Comparison of RRMSE on IEEE (left) and WECC (right) datasets}
\label{fig:RRMSE_combined_base}
\end{figure}

In addition to the previously presented INNs, we also evaluate the \textbf{All-In-One} Model from the FrEIA library, which utilizes a sequential combination of affine coupling layers with soft-clamping and invertible permutations and global affine transformation (ActNorm) to ensure high expressiveness while maintaining exact invertibility.
Overall, the results in Figure \ref{fig:RRMSE_combined_base} show that the class of CF-INNs achieves the most accurate predictions on both training and test trajectories. 
Among them, the affine CF-INNs (All-in-One, Glow, HINT, RealNVP) demonstrate particularly strong generalization capabilities, likely due to the mechanistic structure of coupling layers, which update only a subset of the state dimensions at each step and thereby preserve stability in the learned transformations.
In contrast, NICE as an additive CF-INN exhibits weaker performance on unseen trajectories, while performing well on test trajectories, reflecting the limited expressiveness of additive coupling layers and the resulting difficulty in capturing more complex nonlinear dynamics, an effect that becomes visible on the WECC dataset with its longer observation horizons.
We further observe that the ODE-based INN performs the worst across both datasets and shows limited generalization ability, producing overly smooth mappings that fail to capture the detailed dynamics and thus exhibit limited generalization ability. 
The iResNet also underperforms compared to the CF-INNs, although on the WECC dataset it reaches accuracy levels that are partially comparable to some of the affine coupling models. A possible explanation for this is that the Lipschitz constant overly constrains the learned mapping, and numerical inaccuracies in the fixed-point iteration lead to approximation errors.
On the IEEE 14-bus system dataset, the lowest RRMSE on the training trajectories is obtained by NICE (5.81\%), while RealNVP achieves the best performance on the test trajectories (21.66\%). For the WECC system, the All-In-One model attains the lowest error on the training trajectories (16.77\%), whereas Glow yields the best results on the test data (18.79\%).

\begin{figure}[htbp]
\centering
\includegraphics[width=0.8\linewidth]{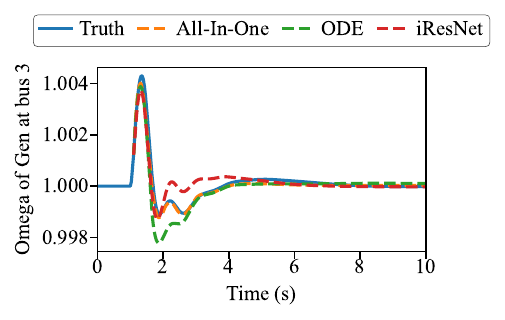}
\caption{Comparison of modeling performance for different INN structures on the IEEE 14-bus system dataset}
\label{fig:CouplingIEEE}
\end{figure}

This observation is further supported by inspecting the trajectory predictions of the individual models.
Figure~\ref{fig:CouplingIEEE} shows a affine CF-INN in contrast to the ODE-based and the iResNet.
The plot reveals that the ODE-based INN and the iResNet struggle to reproduce a clean frequency evolution. 
Both models exhibit noticeable deviations in the predicted trajectories, which aligns with their poor RRMSE performance. Furthermore, we observe that the different classes of INNs introduce different types of errors, which reflects the architectural differences visible in the resulting trajectories.

We further examine the invertibility loss, which is computed solely from the forward and inverse transformations, without involving the Koopman prediction. For CF-INNs, the invertibility error remains at the level of machine precision, as invertibility is guaranteed by architectural design. In contrast, the ODE-based network and iResNet rely on numerical inversion schemes, which leads to higher reconstruction errors on the order of $10^{-9}$ and reflects the degree of practical invertibility of these architectures.

\subsection{Hybrid INNs}
\begin{figure}[htbp]
\centering
\begin{minipage}{0.46\linewidth}
    \centering
    \includegraphics[width=\linewidth]{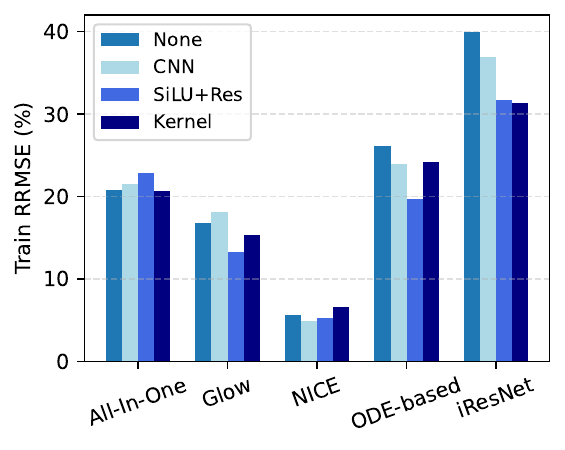}
\end{minipage}
\hfill
\begin{minipage}{0.49\linewidth}
    \centering
    \includegraphics[width=\linewidth]{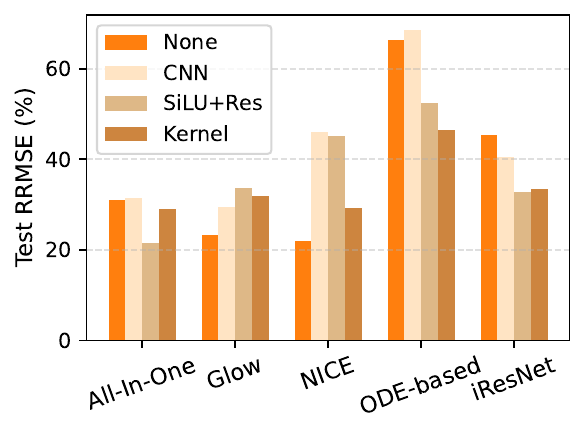}
\end{minipage}
\caption{Comparison of the RRMSE on Training (left) and Test (right) Trajectories on the IEEE-14 bus system}
\label{fig:RRMSE_combined_hybrid}
\end{figure}

To evaluate the effect of the introduced extension networks on model performance, we next analyze the results of the Hybrid INN architectures. Figure~\ref{fig:RRMSE_combined_hybrid} shows the RRMSE of the trajectories for the IEEE 14-bus system. We observe that the CF-INNs do not benefit from the extensions: the affine CF-INNs remain at a similar performance level, while Glow even exhibits a slight degradation on unseen trajectories. NICE, as an additive CF-INN, also performs worse when combined with an extension on the test trajectories. In contrast, the ODE-based INN and iResNet clearly benefit from the extension, achieving improvements of up to 5--10\% in RRMSE. The overall best performance on the training trajectories is achieved by NICE with the CNN extension (4.88\%), whereas on the test trajectories the All-In-One model with the Multitimescale extension performs best (21.63\%).
\label{subsec:hybrid}

We additionally evaluate the best-performing hybrid approaches on the WECC dataset. The All-In-One network shows clear synergy with the Multitimescale extension, achieving slight improvements on the training trajectories and more than a 4\% improvement on the test trajectories, which represents the best overall result in our comparison when considering both training and test performance. A similar improvement had already been observed for the unseen trajectories on the IEEE dataset, suggesting that combining the All-In-One model with the Multitimescale extension leads to enhanced generalization capabilities. In contrast, neither the CNN nor the kernel extension yields improved results.
Moreover, we observe that the performance trends found on the IEEE dataset transfer reasonably well to the WECC dataset. Although the two datasets are not identical, their consistent behavior suggests that our results are not overly sensitive to the specific choice of dataset.

\begin{table}[htbp]
\caption{RRMSE error on training and test trajectories for hybrid INN models for the WECC system.}
\begin{center}
\begin{tabular}{|l|l|c|c|}
\hline
&\multicolumn{3}{|c|}{\textbf{RRMSE [\%]}} \\
\cline{2-4}
\textbf{Extension} & \textbf{\textit{Base Model}}& \textbf{\textit{Train}}& \textbf{\textit{Test}} \\
\hline
 & All-In-One& 16.77 & 23.28 \\
 & Glow & 20.84 & 18.79 \\
- & NICE & 17.76 & 30.35 \\
 & RealNVP & 16.96 & 25.30 \\
 & ODE-based & 32.36 & 43.59 \\
 & iResNet & 36.41 & 33.34 \\
\hline
CNN & All-In-One & 18.24 $\downarrow$ & 26.04 $\downarrow$ \\
\hline
SiLU + Residual & NICE & 14.10 $\uparrow$ & 29.76 $\uparrow$ \\
 & iResNet & 36.65 $\downarrow$ & 32.12 $\uparrow$ \\
& All-In-One& $\mathbf{16.39 \uparrow}$ & $\mathbf{19.57 \uparrow}$ \\
 & Glow & 22.32 $\downarrow$ & 30.93 $\downarrow$ \\
\hline
Kernel & All-In-One & 20.64 $\downarrow$ & 29.13 $\downarrow$ \\
 & NICE & 18.72 $\downarrow$ & 28.20 $\uparrow$ \\
\hline
\end{tabular}
\label{tab:RRMSEComparisonHybridWECC}
\end{center}
\footnotesize
\textit{Note.} $\uparrow$ indicates improvement compared to the base model, while $\downarrow$ indicates degradation.
\end{table}

\begin{figure}[htbp]
\centering
\includegraphics[width=0.8\linewidth]{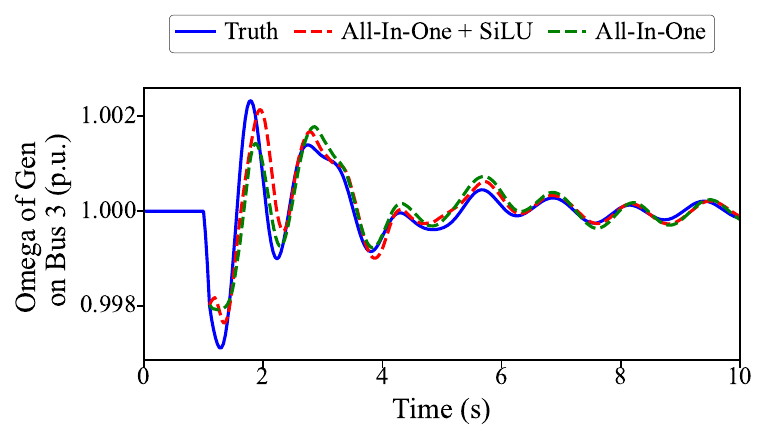}
\caption{Comparison of models on test trajectory 3 on Bus 3 on the WECC dataset}
\label{fig:WECC_Trajectories}
\end{figure}

We further examine these observations by analyzing specific trajectory predictions. Figure~\ref{fig:WECC_Trajectories} illustrates the model performance on a representative test trajectory for the WECC 179-bus system.
In the figure, a comparison is shown between the All-In-One network without any extension and with the SiLU + Residual extension (Multitimescale). In the initial peak of the trajectory, the network combined with the Multitimescale extension tracks the ground truth significantly better, while both networks exhibit similar error on the second peak.

To accurately assess the individual contribution of the non-invertible extension network $\boldsymbol{a}$ in the hybrid architecture, we conduct an ablation study by replacing the complex invertible transformation $\boldsymbol{i}$ with the identity function. This step effectively isolates the effect of the added features $\boldsymbol{z}^a$ on the linear Koopman operator approximation. This configuration, where the latent space is the concatenation of the original state and the extended features, is equivalent to performing Extended Dynamic Mode Decomposition (EDMD) using the extension $\boldsymbol{a}$ as a pre-defined set of Koopman observables. Comparing these results to the full hybrid models allows us to determine if the performance gains originate primarily from the learned invertible feature extraction or the static, expressive feature augmentation.

\begin{table}[htbp]
\caption{RRMSE Comparison of Identity Map (Baseline EDMD) with Feature Extensions}
\begin{center}
\begin{tabular}{|l|c|c|}
\hline
\textbf{Models}&\multicolumn{2}{|c|}{\textbf{RRMSE}} \\
\cline{2-3}
 & \textbf{\textit{Training}}& \textbf{\textit{Test}} \\
\hline
Identity & 41.46 \% & 40.43 \%\\
\hline
Identity + CNN Extension & 36.66 \% & 45.94 \% \\
\hline
Identity + Kernel Extension & 42.69 \% & 44.65 \%\\
\hline
Identity + Multitimescale Extension & 23.03 \% & 31.46 \% \\
\hline
\end{tabular}
\label{tab:ablation_edmd_rrmse}
\end{center}
\end{table}

In Table \ref{tab:ablation_edmd_rrmse}, we can see that the Identity Map, which serves as our benchmark in this case, has a Training RRMSE of $41.46\,$ \% and a Test RRMSE of $40.43\,$ \%. The CNN Extension features provide a seemingly better value on the training trajectories, but ultimately lead to poorer generalization capability, a problem that generally aligns with the observations in Figure \ref{fig:RRMSE_combined_hybrid} and Table \ref{tab:RRMSEComparisonHybridWECC}. The kernel Extension degrades the performance of EDMD on both test and training trajectories, which suggests that the learned Koopman observables are unsuitable. This observation also correlates with the findings regarding the hybrid INNs mentioned above. Only the SiLU + Residual extension significantly improves the RRMSE on both training and test data, suggesting that the learned additional observables accurately reflect the underlying dynamics and can explain the performance increase in combination with networks such as the All-In-One network. In total, we see that the extensions strongly influence the performance of EDMD, both positively and negatively, which motivates their use.

A phenomenon that repeatedly occurs in this study is that models achieve better performance on the test trajectories than on the training data. This behavior can be explained by the structure of the dataset. Each trajectory corresponds to a time-domain simulation of the system’s frequency response to a short 0.1\,s fault applied at a single bus, which results in trajectories that vary in their learning difficulty. Since multiple models are investigated, it is therefore of interest to analyze which trajectories are best learned by which model and whether model extensions alter these properties.

\begin{figure}[htbp]
\centering
\begin{minipage}{0.46\linewidth}
    \centering
    \includegraphics[width=\linewidth]{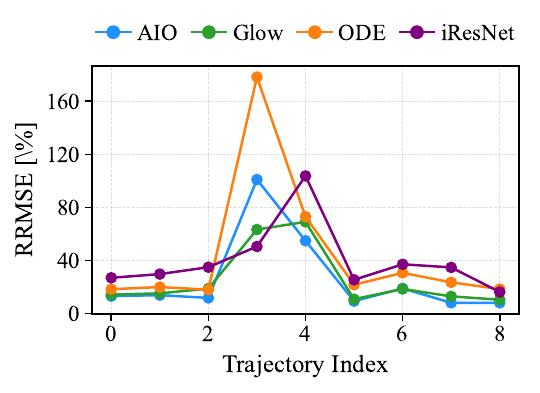}
\end{minipage}
\hfill
\begin{minipage}{0.51\linewidth}
    \centering
    \includegraphics[width=\linewidth]{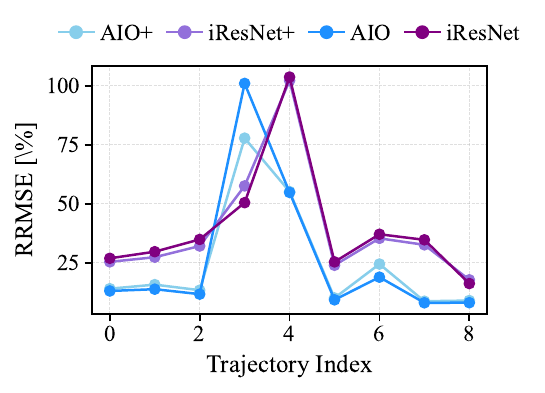}
\end{minipage}

\caption{RRMSE per trajectory of test trajectories for base INNs (left) and with Multitimescale extension (right) on the WECC dataset. Models marked with '+' denote the hybrid variant incorporating the proposed Multitimescale extension layer.}
\label{fig:rrmse_per_traj}
\end{figure}

In Figure \ref{fig:rrmse_per_traj}, we observe that the INNs encounter particularly strong difficulties with a few specific trajectories. On the left, it can be seen that the Glow and iResNet networks exhibit a different structure compared to the other INNs. On the right, we notice that the extension for iResNet hardly affects the learning properties of the trajectories, whereas the All-In-One extension significantly improves the learning of trajectory 3 but slightly worsens trajectory 6. Test trajectory 3, also shown in Figure \ref{fig:WECC_Trajectories}, displays a distinctly irregular structure with many small fluctuations, which poses challenges for the models, but can be improved by the Multitimescale extension. This further illustrates how the application of extensions can alter the characteristics of model behavior.

\section{Discussion}
\label{sec:Disc}
This work investigated the suitability and synergy of various INN architectures and their hybrid extensions within a Koopman operator framework for power system dynamics. We find that prediction accuracy and generalization are highly dependent on the INN architecture. CF-INNs consistently achieved the lowest error values and strongest generalization due to the inherent structural stability. The integration of non-invertible feature extensions into the hybrid architecture revealed a highly architecture-specific synergy, substantially benefiting weaker base models like the ODE-based INN and iResNet (improving RRMSE by up to $10\%$), but failing to yield gains for the already strong CF-INNs. The All-In-One model combined with the Multitimescale extension achieved the best overall result on the challenging WECC test set ($19.57\%$), confirming that optimal performance relies on identifying the right hybrid combination in the specific use case. The combination of the feature space extensions with the identity map provides insights into which extensions deliver effective features for Koopman Lifting. This establishes a good a-priori evaluation method for assessing the suitability of an extension and enables a targeted selection of promising architectures.

\section{Conclusion}
\label{sec:conc}
This study concludes that Affine Coupling INNs performed best consistently, offering an optimal balance between structural stability and expressive power. Thus, INNs using the Koopman approach enable accurate trajectory prediction without the need for a model-based simulation tool.
While non-invertible feature extensions can enhance the performance of weaker base models, such as the ODE-based INN and iResNet, they may also provide improvements for some already strong models, such as All-In-One. However, we also observed cases where extensions degraded performance. This demonstrates that the benefits of extensions are highly architecture- and dataset-dependent, requiring careful evaluation for each application.


\begin{thebibliography}{00}
\bibitem{b1} Hosseinzadeh, Nasser, et al. "Voltage stability of power systems with renewable-energy inverter-based generators: A review." Electronics 10.2 (2021): 115.
\bibitem{b2} N. Hatziargyriou et al., "Definition and Classification of Power System Stability – Revisited \& Extended," IEEE Transactions on Power Systems 36.4 (2021): 3271-3281.
\bibitem{b3} Schmid, Peter J. "Dynamic mode decomposition and its variants." Annual Review of Fluid Mechanics 54.1 (2022): 225-254.
\bibitem{b4} Lusch, Bethany, J. Nathan Kutz, and Steven L. Brunton. "Deep learning for universal linear embeddings of nonlinear dynamics." Nature communications 9.1 (2018): 4950.
\bibitem{b5} Takeishi, Naoya, Yoshinobu Kawahara, and Takehisa Yairi. "Learning Koopman invariant subspaces for dynamic mode decomposition." Advances in neural information processing systems 30 (2017)
\bibitem{b6} Mezić, Igor. "Spectral properties of dynamical systems, model reduction and decompositions." Nonlinear Dynamics 41.1 (2005): 309-325.
\bibitem{b7}
Frion, A., Drumetz, L., Mura, M. D., Tochon, G., \& Aïssa-El-Bey, A. (2025). Augmented Invertible Koopman Autoencoder for long-term time series forecasting. arXiv preprint arXiv:2503.12930.
\bibitem{b8} Dinh, Laurent, David Krueger, and Yoshua Bengio. "Nice: Non-linear independent components estimation." arXiv preprint arXiv:1410.8516 (2014).
\bibitem{b9} Dinh, Laurent, Jascha Sohl-Dickstein, and Samy Bengio. "Density estimation using real nvp." arXiv preprint arXiv:1605.08803 (2016).
\bibitem{b10} Kingma, Durk P., and Prafulla Dhariwal. "Glow: Generative flow with invertible 1x1 convolutions." Advances in neural information processing systems 31 (2018).
\bibitem{b11} Kruse, Jakob, et al. "HINT: Hierarchical invertible neural transport for density estimation and Bayesian inference." Proceedings of the AAAI Conference on Artificial Intelligence. Vol. 35. No. 9. 2021.
\bibitem{b12} Jacobsen, Jörn-Henrik, Arnold Smeulders, and Edouard Oyallon. "i-revnet: Deep invertible networks." arXiv preprint arXiv:1802.07088 (2018).
\bibitem{b13} Gomez, Aidan N., et al. "The reversible residual network: Backpropagation without storing activations." Advances in neural information processing systems 30 (2017).
\bibitem{b14} Behrmann, J., Grathwohl, W., Chen, R. T., Duvenaud, D., \& Jacobsen, J. H. (2019, May). Invertible residual networks. In International conference on machine learning (pp. 573-582). PMLR.
\bibitem{b15}
Behrmann, J., Vicol, P., Wang, K. C., Grosse, R., \& Jacobsen, J. H. (2021, March). Understanding and mitigating exploding inverses in invertible neural networks. In International Conference on Artificial Intelligence and Statistics (pp. 1792-1800). PMLR.
\bibitem{b16}
Jin, Y., Hou, L., Zhong, S., Yi, H., \& Chen, Y. (2023). Invertible Koopman network and its application in data-driven modeling for dynamic systems. Mechanical Systems and Signal Processing, 200, 110604.
\bibitem{b17}
Jin, Yuhong, Lei Hou, and Shun Zhong. "Extended dynamic mode decomposition with invertible dictionary learning." Neural networks 173 (2024): 106177.
\bibitem{b18}
Chen, R. T., Rubanova, Y., Bettencourt, J.,\& Duvenaud, D. K. (2018). Neural ordinary differential equations. Advances in neural information processing systems, 31.
\bibitem{b19}
Grathwohl, W., Chen, R. T., Bettencourt, J., Sutskever, I., \& Duvenaud, D. (2018). Ffjord: Free-form continuous dynamics for scalable reversible generative models. arXiv preprint arXiv:1810.01367.
\bibitem{b20}
Meng, Y., Huang, J., \& Qiu, Y. (2024). Koopman operator learning using invertible neural networks. Journal of Computational Physics, 501, 112795.
\bibitem{b21} 
Li, Xiao, Xinyi Wen, and Benjamin Schäfer. "Learning the Frequency Dynamics of the Power System Using Higher-order Dynamic Mode Decomposition." arXiv preprint arXiv:2502.06186 (2025).
\bibitem{b22} 
H. Cui, F. Li, and K. Tomsovic, “Hybrid Symbolic-Numeric Framework
for Power System Modeling and Analysis,” IEEE Trans. Power Syst.,
vol. 36, no. 2, pp. 1373–1384, Mar. 2021.
\bibitem{b23} 
Lynton Ardizzone, Till Bungert, Felix Draxler, Ullrich Köthe, Jakob Kruse, Robert Schmier, and Peter Sorrenson.
Framework for Easily Invertible Architectures (FrEIA), 2018-2022.
\bibitem{b24} 
Vega, Jose Manuel, and Soledad Le Clainche. Higher order dynamic mode decomposition and its applications. Academic Press, 2020.



\end{thebibliography}
\end{document}